\documentclass[preprint,12pt,numbers]{elsarticle}

\usepackage{amssymb}

\usepackage{amsmath}

\usepackage[a4paper, total={6in, 10in}]{geometry}

\usepackage{ulem}
\usepackage{cancel}

\journal{Journal of Computational Physics}
\usepackage[colorlinks=true,linkcolor=blue,citecolor=blue,urlcolor=blue]{hyperref}

\newcommand{\dbd}[2]{\frac{\partial#1}{\partial#2}}
 \newcommand{\dbdt}[2]{\frac{\partial^2#1}{\partial#2^2}}

\newcommand{\bmat}[1]{ \begin{bmatrix} #1 \end{bmatrix}}

\newcommand{\sRe}{_{\rm re}} \newcommand{\sIm}{_{\rm im}}
\newcommand{\half}{\frac{1}{2}}
\newcommand{\mcL}{\mathcal{L}}

\begin{document}

\begin{frontmatter}

\title{ Fast Synergetic Simulation to Study Slow Evolution of Soliton Patterns in Optical Resonators }


\author[label1]{Sanzida Akter}
\author[label1]{Pradyoth Shandilya}
\author[label1]{Logan Courtright}
\author[label2]{Amir Leshem}
\author[label1]{Giuseppe D’Aguanno}
\author[label2]{Omri Gat}
\author[label1]{Curtis R. Menyuk}


\affiliation[label1]{
    organization={University of Maryland, Baltimore County},
    addressline={1000 Hilltop Circle},
    city={Baltimore},
    state={Maryland},
    postcode={21250},
    country={USA}
}

\affiliation[label2]{
    organization={Racah Institute of Physics, The Hebrew University},
    city={Jerusalem},
    postcode={9190401},
    country={Israel}
}


\begin{abstract}

In physical systems, including biological systems, complex patterns often emerge gradually over long timescales through slow collective dynamics controlled by a few key variables. In particular, optical patterns arise as ordered spatiotemporal structures due to nonlinear light-matter interactions, analogous to morphogenesis observed in biological systems. In a driven nonlinear Kerr cavity, dissipative Kerr solitons are an important example of optical pattern formation, arising through the self-organization of coherent localized structures that generate optical frequency combs and have a wide range of practical applications. When multiple solitons coexist in a nonlinear cavity with sufficiently large separations, their interactions can evolve over timescales far exceeding the characteristic loss and gain timescales, serving as an example of slow pattern formation. Numerically modeling these slow interaction dynamics has long been challenging due to step size limitations imposed by stiffness in the governing equations due to the necessity of accurately resolving a large number of rapidly damped degrees-of-freedom. In this work, we present a numerical scheme called the synergetic method, which efficiently models slow processes by eliminating rapidly damped degrees-of-freedom and thereby enabling step sizes many orders of magnitude larger than what is possible using conventional methods. We apply our method to study the slow interactions of complex soliton molecules evolving on laboratory timescales, demonstrating its ability to operate $10^3$ to $10^5$ times faster than conventional numerical approaches. Our method has made it possible to model the full interaction dynamics of a three-soliton molecule, and we also apply it to study of evolution of an eight-soliton molecule. We expect that this approach will be of use in the study of pattern formation across a broad array of physical systems.

\end{abstract}



\end{frontmatter}


\section{Introduction}

Nonlinear systems far from equilibrium often exhibit spontaneous pattern formation, a phenomenon observed across diverse fields. Turing first showed that an initially uniform chemical medium can self-organize into stable spatial patterns through a diffusion-driven instability, providing a chemical basis for morphogenesis \cite{turing_1990}. A few years later, Hermann Haken reformulated this idea in general terms as part of his book titled \textit{Synergetics: An Introduction} \cite{haken_1977}. He introduced the concept of order parameters to denote the small set of collective modes that dominate a system’s behavior near an instability, while labeling the remaining fast-relaxing degrees-of-freedom (DOFs) as slaved variables.  In essence, the long-term behavior of a complex system is governed by a handful of slow collective DOFs, whereas all other microscopic variables quickly relax and adjust to follow the pattern set by those dominant modes. Self-organization causes an enormous reduction in effective DOFs, replacing microscopic chaos with robust large-scale order.  In developmental biology, reaction-diffusion mechanisms of the Turing type are now thought to underlie many spontaneous patterns, such as stripes on zebras and spots on leopards, the ridges of human fingerprints, and the pigmentation of flower petals \cite{hunter_2023,vittadello_2021}. In fluid systems, a classic example is Rayleigh–Bénard convection, where the symmetry of a uniformly heated fluid layer spontaneously breaks to form a regular array of convection cells once a critical temperature gradient is exceeded \cite{ahlers_2009}. Similarly, in laser physics, when a laser medium is pumped above threshold, a single optical mode---the coherent electromagnetic field---grows to dominate the behavior of countless microscopic emitters. The collective laser field thus acts as an order parameter that enslaves the individual atomic dipole oscillations, forcing them into phase and producing a highly coherent beam \cite{haken_1977}.

Optical systems are a rich source of spontaneous pattern formation, mirroring the self-organization seen in chemical and biological systems. Just as the symmetry of a reaction-diffusion medium can be broken to obtain spatial patterns, a nonlinear optical cavity can likewise destabilize a uniform light field and develop structured intensity patterns \cite{lugiato_1987,mcdonald_1990,haelterman_1992}. One particular example, of great contemporary interest due to its applications to compact precision measurement \cite{kippenberg_2018}, is a Kerr microresonator pumped with a continuous wave (CW) laser through an input waveguide. Since the 2007 discovery by Del'Haye et al.~\cite{delhaye_2007} that microresonators can generate frequency combs, these devices have been widely used in numerous applications, including telecommunications, spectroscopy, metrology, and LiDAR \cite{kippenberg_2011, sun_2023, diddams_2020}.

\begin{figure}[htb]
\centering
\includegraphics[width=1\textwidth]{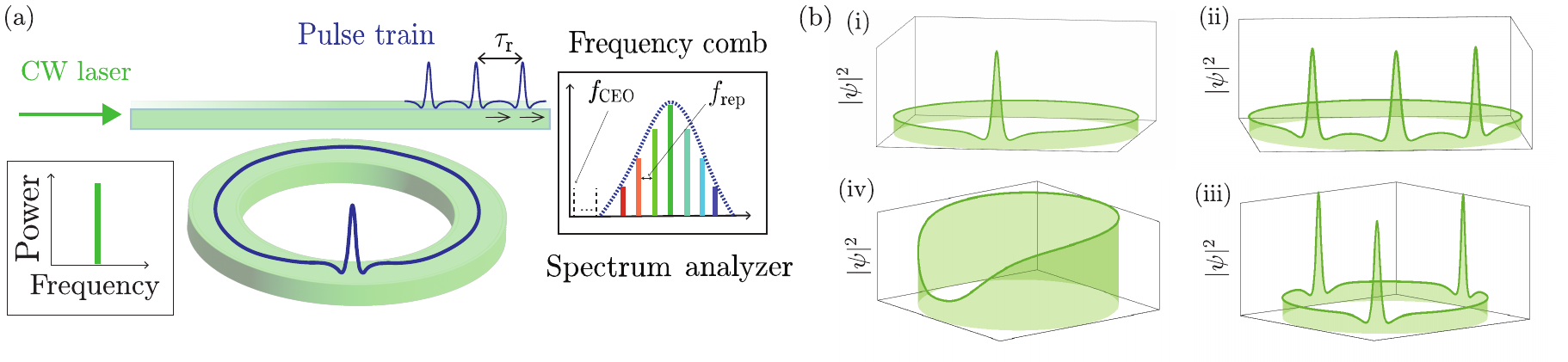} 
\vspace{-15pt}
\caption{(a) Schematic of a microresonator system for frequency comb generation. A microring resonator is driven by an external CW laser source to generate a train of soliton pulses or more complex patterns that form a frequency comb, as analyzed using an optical spectrum analyzer. (b) Intracavity intensity profile illustration of commonly observed optical patterns. Anomalous dispersion regime: (i) a single dissipative Kerr soliton, (ii) a soliton molecule, (iii) a perfect soliton crystal. Normal dispersion regime: (iv) a dark soliton. }
\label{fig1}
\end{figure}

Shortly after the discovery of microresonator frequency combs, it was demonstrated that they can be modeled using the Lugiato-Lefever equation (LLE)---the same model equation that Lugiato and Lefever had used earlier to describe an optical cavity \cite{lugiato_1987,chembo_2013}.  This model, which is also referred to as the driven-damped nonlinear Schr\"odinger equation, shows how the interplay of dispersion, Kerr nonlinearity, cavity loss, and an external pump can act together to destabilize the homogeneous steady state and produce optical patterns. Essentially, an initially flat intracavity field induced by the CW laser undergoes a Turing-like periodic instability, and once the four driving terms mutually balance, these self-organized optical patterns appear as stationary, localized structures known as solitons. Upon coupling from the microresonator into the output waveguide, these precisely spaced circulating solitons emerge as a periodic pulse train, which corresponds to an optical frequency comb in the spectral domain [Fig.~\ref{fig1}(a)].

Microresonators support different types of soliton patterns depending on the dispersion regime and external pump conditions as described in Fig.~\ref{fig1}(b) \cite{zhen_19, xue_2015}, where we show a single bright soliton, soliton molecules, a dark soliton or platicon, and a soliton crystal \cite{karpov_2019, lobanov_2015, stratmann_2005}. In particular, soliton molecules [Fig.~\ref{fig1}(b)-(ii)] are stable bound states in which two or more solitons interact through a balance of attractive and repulsive forces, settling into a robust equilibrium configuration. The difference between soliton molecules and soliton crystals [Fig.~\ref{fig1}(b)-(iii)] is that soliton molecules consist of solitons bound together at a fixed distance apart that is unrelated to the resonator circumference, whereas soliton crystals are periodic arrangements of solitons that are evenly spaced within the resonator. 

The formation, interaction, and collision dynamics of soliton molecules offer insights into general principles of solitary wave interactions and emergent structures in nonlinear systems, analogous to molecular and biological self-organization \cite{weng_2020}. From an engineering standpoint, soliton molecules significantly expand the potential of frequency comb technology. The spacing of solitons within these molecules provides additional DOFs, enabling the
possibility of transferring optical data and surpassing the limitation of binary coding \cite{akhmediev_1994}. Soliton molecule systems have long been the subject of both theoretical \cite{barashenkov_1998,pedro_2018, herr_2014} and experimental \cite{stratmann_2005, krupa_2017, peng_2018} interest.

In soliton molecules, as the separation between individual solitons increases, the interaction strength decreases exponentially \cite{pedro_2018, leshem_2026}, resulting in a slow evolution of complex solitary waves and the emergence of equilibrium states over extended timescales analogous to natural pattern formation. Due to numerical step size limitations arising from the inherent stiffness \cite{gear_1971,hairer_1996,recipes_2007} of the system, which in turn is due to the large difference in timescales between the dissipative timescale and the timescale on which well-separated solitons interact, accurately modeling these slowly evolving processes is computationally challenging, despite their widespread occurrence and considerable scientific interest.

These slow, long-range soliton interactions have been observed in a variety of experimental platforms. In driven fiber ring resonators, ultraweak coupling between temporal cavity solitons has been observed over effective propagation distances reaching astronomical scales, mediated by transverse acoustic waves excited through electrostriction \cite{jang_ultraweak}.  In passively mode-locked fiber lasers, elastic soliton crystals with tunable bond lengths have been shown to assemble and reorganize over seconds of laboratory time \cite{andrianov_2021}, and breather molecular complexes spanning hundreds of picoseconds have been bound by weak dispersive-wave exchange \cite{peng_2021}. Other studies have documented gain-modulation-driven oscillations between molecular and dissociated soliton states \cite{zhou_2021},  and controlled merging or annihilation of localized dissipative structures in driven Kerr resonators \cite{jang_2016}. A common thread runs through all of these works: The dominant interactions are weak, non-local, and act over timescales many orders of magnitude longer than the round-trip or photon-lifetime timescales of the underlying equations, making direct numerical verification computationally infeasible with conventional methods. Numerical studies that include delayed feedback or higher-order dispersion further suggest that many regimes of long-range organization predicted by theory remain experimentally unverified for the same reason \cite{frank_2023}.

In this work, we propose a numerical method specifically designed for modeling slow processes commonly observed in nature that can be modeled using nonlinear partial differential equations, and we apply this approach to study the slow evolution and formation of complex soliton molecules in microresonator systems. While we previously presented a brief description of this method and some results in conference abstracts, this paper is the first complete description of the method \cite{synergetic_2024, akter_2025_noise, synergetic_2025}.  Inspired by Haken’s framework \cite{haken_1977}, we name our method the \textit{synergetic method}, highlighting its capability to efficiently capture the slow dynamics governed by a small number of dominant DOFs. We first validate our method by modeling the interaction dynamics of a two-soliton system for which we previously derived an analytical expression for the potential function \cite{leshem_2026}. After validation, we next use the synergetic method to investigate and map the two-dimensional potential function of a three-soliton molecule, a result for which analytical solutions are not available and would be extremely time-consuming to obtain through conventional simulations. Next, we investigate the slow interaction dynamics of an eight-soliton molecule with our method, tracking its evolution for nearly $100$ seconds of laboratory time, representing the first LLE simulation to span such extended durations with any numerical technique. Throughout this work, we demonstrate the versatility of the synergetic method by applying it to several soliton molecule configurations, thereby establishing its suitability for exploring analogous pattern formation phenomena across diverse natural systems. 

\section{Basic Equations and the Synergetic Method}

\subsection{Governing equations}
The microresonator system that we will consider here is one of the numerous nonlinear physical systems that can be described at lowest order by the nonlinear Schr\"odinger equation \cite{lugiato_1987, chembo_2013}. We begin by considering the propagation of an electric field inside a high-$Q$ optical cavity in a single transverse mode with an external pump field and loss that can be described by a driven-damped generalized nonlinear Schr\"odinger equation \cite{lugiato_1987, coen_2012, chembo_2010},
\begin{equation} \begin{split}
    \frac{\partial a(\theta,\tau)}{\partial\tau} &= -\frac{\kappa}{2}a(\theta,\tau) + i{\cal D}(\theta)*a(\theta,\tau) - i\gamma|a(\theta,\tau)|^2a(\theta,\tau) \\ &\hspace{1cm}+\sqrt{\kappa_{\rm e}}E_{\rm wg}\exp(im_0\theta + i\omega_{\rm wg}t), \label{Eq1:M1}
 \end{split}\end{equation}
where $a(\theta,\tau)$ is the electric field envelope as a function of the azimuthal coordinate $\theta$ and time $\tau$, normalized so that $(1/2\pi)\int_0^{2\pi}|a(\theta,\tau)|^2\,d\theta$ is the total energy in the cavity, $\kappa$ is the cavity loss coefficient, ${\cal D}(\theta)$ is the dispersion operator, $\gamma$ is the Kerr coefficient, $\kappa_{\rm e}$ is the coupling coefficient for the driving power from the waveguide, and $E_{\rm wg}$ is the waveguide power amplitude, normalized so that $|E_{\rm wg}|^2 = P_{\rm wg}$ is the input waveguide power. The mode number of the waveguide resonance that is pumped is $m_0$, and the radial frequency of the pump is $\omega_{\rm wg}$. Defining the transform pair between the azimuthal and mode number domains
 \begin{equation}
     A_m(\tau) = \frac{1}{2\pi}\int_{-\pi}^{\pi} a(\theta,\tau)\exp(-im\theta)\,d\theta, \quad
     a(\theta,\tau) = \sum_{m=-\infty}^\infty A_m(\tau)\exp(im\theta)\,,
 \end{equation}
we can write the dispersion operator as
 \begin{equation}
     {\cal D}(\theta)*a(\theta,\tau) = \sum_m \omega_m A_m(\tau)\exp(im\theta),
 \end{equation}
where $\omega_m$ is the radial oscillation frequency of the $m$-th mode.  In order to drive a high-$Q$ optical cavity, the waveguide mode that is pumping the resonator must have a frequency that is close to the frequency of one of the resonances $m=m_0$. We will focus on systems for which the bandwidth of the optical waveform in the cavity is sufficiently narrow that we can approximate the dispersion $\omega_m$ by a  second order polynomial expansion in $\mu = m-m_0$, so that
 \begin{equation}
     \omega_{m-m_0} = \omega_0 + D_1\mu + \frac{1}{2}D_2 \mu^2.
 \end{equation}
To obtain the LLE from Eq.~\eqref{Eq1:M1}, we first transform the field, letting $\bar a(\theta,\tau) = a(\theta,\tau)$ $\exp(-im_0\theta - i\omega_{\rm wg}\tau)$, and we then transform the azimuthal coordinate so that $\bar\theta = \theta + D_1\tau$.  We then find
 \begin{equation} \begin{split}
     \dbd{\bar a(\bar\theta,\tau)}{\tau} &= \left(-\frac{\kappa}{2} +i\sigma \right)\bar a(\bar\theta,\tau)  -i\frac{D_2}{2} \dbdt{\bar a(\bar\theta,\tau)}{\bar\theta} -i\gamma|\bar a(\bar\theta,\tau)|^2\bar a(\bar\theta,\tau) +\sqrt{\kappa_{\rm e}}E_{\rm wg}. \label{Eq5:M6a}
 \end{split}\end{equation}
where $\sigma = \omega_0 - \omega_{\rm wg}$ is the difference between the resonant and pump frequencies.  The first transformation has the effect of removing the rapid temporal and azimuthal oscillations of the pump field, while the second transformation has the effect of removing the angular group velocity motion.  We now take the complex conjugate of Eq.~\eqref{Eq5:M6a} and normalize times with respect to the photon lifetime $\tau_{\rm ph} = 2/\kappa$, waveform energy with respect to the Kerr coefficient $\gamma$, and the azimuthal coordinate with respect to the dispersive scale length, so that $t = \kappa \tau/2$, $\alpha = 2\sigma/\kappa$, $x=(\kappa/2D_2)^{1/2}\bar\theta$, $\psi = (2\gamma/\kappa)^{1/2}\bar a^*$, and $F= (2/\kappa)^{3/2}(\gamma\kappa_{\rm e})^{1/2}E_{\rm wg}$. We have $-X/2\le x<X/2$, where $X=(\kappa/2D_2)^{1/2}2\pi$.  Letting $\partial\psi(x,t)/\partial t = f[\psi(x,t)]$, we thus obtain
 \begin{equation}
     \dbd{\psi(x,t)}{t} = f[\psi(x,t)] = -(1+i\alpha)\psi(x,t) + \frac{i}{2}\dbdt{\psi(x,t)}{x} + i|\psi(x,t)|^2\psi(x,t) + F. \label{Eq6:LLE}
 \end{equation}

While the details of the transformation that brought us from Eq.~\ref{Eq1:M1} to Eq.~\ref{Eq6:LLE} are particular to the microresonator, a similar transformation appears in the many physical systems in which the nonlinear Schr\"odinger equation appears as the zero-order model.

\subsection{Numerical timescale of the LLE}
The dynamical timescale of the LLE is defined in terms of the photon lifetime, which typically falls within the nanosecond range in modern microresonators with high $Q$-factors \cite{yang2018,lee_2012,armani_2003}. With the chosen normalization in Eq.~\ref{Eq6:LLE}, a time step of $\Delta t=1$ in simulation corresponds to an evolution over one photon lifetime in the resonator, while typical step sizes range from $10^{-3}\tau_{\mathrm{ph}}$ to $10^{-2}\tau_{\mathrm{ph}}$ to avoid numerical instabilities and retain accuracy. Given the step sizes achievable with existing methods, it is common to typically simulate only short time segments to validate experimentally observed processes that unfold over timescales ranging from seconds to hours \cite{jang_ultraweak}.

As a concrete example, we consider the microresonator system described by Yang et al.~\cite{yang2018}, characterized by a quality factor $Q \approx 2 \times 10^8$ with $\tau_{\rm{ph}} = 340\text{ ns}$. Given the rapidly damped radiation modes in the LLE \cite{zhen_19}, the time step $\Delta t$ is constrained to a range between 0.34 ns ($10^{-3}\tau_{\rm{ph}}$) and 3.4 ns ($10^{-2}\tau_{\rm{ph}}$), depending on the rate of change of $\psi(x,t)$. Utilizing the conventional split-step Fourier method (SSFM) \cite{agarwal_nfo}, the simulation of an evolution spanning one photon lifetime requires approximately 0.08 s of computational time on a high-performance workstation ($13$th Gen Intel Core i7, 32GB memory). Consequently, systems exhibiting evolution over the microsecond to millisecond range can be efficiently modeled using the SSFM with computational durations ranging from 0.3 to 300 s, representing a notably rapid processing capability. However, the same method proves inefficient as $\psi(x, t)$ begins to evolve gradually over timescales spanning seconds to hours, necessitating computational durations of $2.353 \times 10^5 \, \text{s}$ (approximately 65 hours) to $2.353 \times 10^8 \, \text{s}$ (approximately $6.5 \times 10^4$ hours), making simulations of evolution on these timescales impractical.

A fundamental challenge in simulating stiff systems like the LLE is the necessity to resolve the timescale of the most rapidly evolving DOFs, even when those DOFs are quickly damped. Consequently, the distinct separation of timescales between the few relevant DOFs and the negligible ones leads to highly stiff governing equations. This stiffness is strongly evident in the LLE, where solitons, which are characterized by a small number of DOFs, interact on timescales that are many orders of magnitude longer than the photon lifetime.\\

\subsection{The Synergetic Method}
Since the system that we are considering is autonomous, the change in $\psi$ may be written through lowest order in $\Delta t$, a change in time with respect to an initial time $t_0$, as
 \begin{equation}
     \dbd{\Delta\psi(x,t)}{t} = f[\psi(x,t_0)] + \mathcal{L}[\psi(x,t_0)]\Delta\psi(x,t), \label{Eq7:FoEvEq}
 \end{equation} 
where $\mathcal{L}$ is the Jacobian of $f$.  Since Eq.~\ref{Eq6:LLE} includes both $\psi$ and $\psi^*$ in the nonlinear term, it is necessary to treat these quantities as independent when evaluating the Jacobian.  In analytical work, it is common to keep $\psi$ and $\psi^*$ as the independent variables \cite{lamb_1980,ablowitz_1981,hasegawa_1995,kaup_1990,georges_1995}, but for computational work, it is more convenient to divide Eq.~\ref{Eq6:LLE} into its real and imaginary parts \cite{Wang_2014}.  Writing $\psi = \psi\sRe + i\psi\sIm$ and letting ${\Psi} = [\psi\sRe\>\> \psi\sIm]^{\rm T}$, we obtain
 \begin{equation} 
     \dbd{{\Psi}}{t} = {f}_{\Psi}(\psi\sRe,\psi\sIm) = \bmat{f_{\rm re} \\ f_{\rm im}} = {\renewcommand{\arraystretch}{1.6} \bmat{-\psi\sRe + \alpha\psi\sIm - \half\dbdt{\psi\sIm}{x}-\gamma\left( \psi\sRe^2 + \psi\sIm^2\right)\psi\sIm + F \\ 
         -\psi\sIm  - \alpha\psi\sRe + \half\dbdt{\psi\sRe}{x} + \gamma\left( \psi\sRe^2 + \psi\sIm^2\right)\psi\sRe} }.
         \label{Eq8:RLLE}
 \end{equation}
The Jacobian now becomes
 \begin{equation}
\renewcommand{\arraystretch}{1.8}
  \mathcal{L} = \begin{bmatrix}
        \frac{\partial f\sRe}{\partial \psi_{\mathrm{re}}} & \frac{\partial f\sRe}{\partial \psi_{\mathrm{im}}} 
        \\ \frac{\partial f\sIm}{\partial \psi_{\mathrm{re}}} & \frac{\partial f\sIm}{\partial \psi_{\mathrm{im}}}
    \end{bmatrix}
    = \begin{bmatrix}
        -\mathcal{I}-2\psi_{\mathrm{re}} \psi_{\mathrm{im}} & \alpha - \frac{1}{2}\frac{\partial ^2 }{\partial x^2} - \psi_{\mathrm{re}}^2 -3 \psi_{\mathrm{im}}^2 \\ -\alpha + \frac{1}{2}\frac{\partial ^2 }{\partial x^2} + \psi_{\mathrm{re}}^2 +3 \psi_{\mathrm{im}}^2 & 
        -\mathcal{I}+2\psi_{\mathrm{re}} \psi_{\mathrm{im}}
     \end{bmatrix},
        \label{Eq9:define_L}
\end{equation}
where $\mathcal{I}$ is the identity operator. From the structure of $\mathcal{L}$, it follows that the eigenvalues are real or must come in complex conjugate pairs with adjoint eigenfunctions.  If the real parts of any of the eigenvalues are greater than zero, then the solution is linearly unstable.
\begin{figure}[httb]
\centering
\includegraphics[width=1\textwidth]{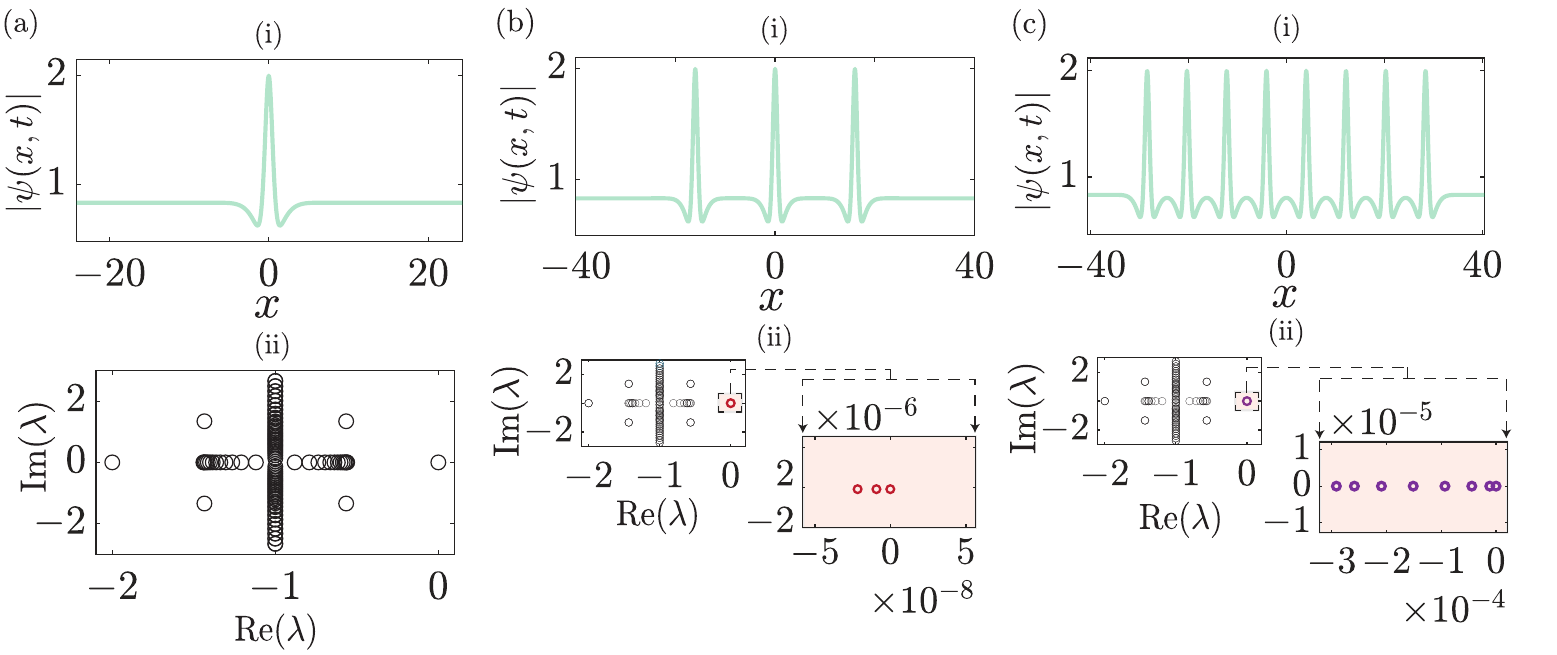} 
\vspace{-10pt}
\caption{(a) (i) Profile of a single soliton stationary solution. (ii) The corresponding eigenvalue spectrum derived from the linearized operator. The spectrum reveals a fourfold symmetry and a single eigenvalue near zero, corresponding to translational invariance of the soliton. (b) (i) A three-soliton molecule solution and (ii) its eigenvalue profile. Here, there are three eigenvalues near zero: one represents translational invariance (similar to the single soliton), while the other two represent the slowly varying DOFs associated with the relative motion between solitons. (c) (i) An eight-soliton molecule and (ii) its eigenvalue spectrum. In this system, there are eight eigenvalues near zero, where one corresponds to translational invariance and the remaining seven refer to relative motion.}
\label{molecule_eigenvalues}
\end{figure} \\

In the case that $f[\psi] = 0$, the solution is stationary, and the eigenvalue spectrum of the operator $\mathcal{L}$, also referred to as the dynamical spectrum, determines the linear stability of this solution and its response to perturbations \cite{menyuk_2016}.  This approach has been used to determine the stability of cnoidal waves \cite{zhen_19} and the response of single solitons to thermorefractive noise \cite{moille_2025} in microresonators.  It has also been used to study the stability of soliton molecules \cite{barashenkov_1998,pedro_2018, akter_2025_noise}.  In Fig.~\ref{molecule_eigenvalues}, we show the dynamical spectrum for three stationary solutions:  a single soliton [Fig.~\ref{molecule_eigenvalues}(a)], a three-soliton molecule [Fig.~\ref{molecule_eigenvalues}(b)], and an eight-soliton molecule [Fig.~\ref{molecule_eigenvalues}(c)]. All dynamical spectra have a characteristic four-fold symmetry that is explained in \cite[Supplement]{zhen_19}, and an eigenvalue $\lambda$ at $\lambda=0$, which is a consequence of the translational invariance of the LLE\@.  Finally, we observe that all but a few of the eigenvalues have negative real parts that are within a factor of two of $-1$ and correspond to modes that damp during a photon lifetime.  However, in the case of the three-soliton and eight-soliton molecules  [Fig.~\ref{molecule_eigenvalues}(b), (c)], we observe a cluster of real eigenvalues whose magnitudes are four orders of magnitude smaller than $-1$. The  eigenmodes correspond physically to mutual soliton interactions.  The numerical approach that we used to find the stationary solution is described in \cite{Wang_2014,zhen_19}.  It makes use of the Levenberg-Marquardt algorithm as implemented in MATLAB's {\tt FSOLVE} \cite{MATLAB_fsolve} routine.

The synergetic method is based on the observation that when quasi-stationary multi-soliton patterns form, the dynamical spectrum of the Jacobian $\mathcal{L}$ is similar to the spectrum of the stationary solutions and has a small cluster of eigenvalues governing the mutual soliton interactions whose magnitudes are many orders of magnitude lower than the rest of the eigenvalues, although we note that their real parts can be positive as well as negative.  We therefore only keep the components of $f$ that consist of the weakly damped or growing eigenmodes, which makes it possible to take time steps that are many orders of magnitude larger than the photon lifetime.

Numerical approaches for solving differential equations rely on repeated use of the equation of motion, which in our case is Eq.~\ref{Eq6:LLE}.  That is also the case for traditional stiff solvers \cite{gear_1971,hairer_1996,recipes_2007}.  In our synergetic approach, we first run a standard simulation for typically $ 100 \tau_{\rm{ph}}$ to ensure that any transients have dissipated, and we have arrived at a quasi-stationary solution. We then start with the first-order approximation as described in Eq.~\ref{Eq7:FoEvEq}.  The operator $\mcL$ has a small number of eigenvalues clustered near $\lambda =0$ for multi-soliton systems, but are not equal to zero, that define an $m$-dimensional subspace of eigenvectors $v_j$, such that $\mcL v_j = \lambda_jv_j$, along with the adjoint eigenvectors, $w_j$, that satisfy $\mcL^{\dagger}w_j=\lambda^{*}_jw_{j}$ so that the following normalization holds
\begin{equation} \label{Eq10:AdjDef}
\frac{1}{X}\int_{-X/2}^{X/2} w_j^{\dagger} v_k\,dx = \delta_{jk},
\end{equation}\\
and so that the adjoint vectors are orthogonal to not just the $v_k$ ($k\ne j$) in the $m$-dimensional subspace with eigenvalues close to zero, but to all the eigenvectors in the spectrum of $\mcL$ except $v_j$.  We next define $f_{\rm syn}[\psi(t)] = \mathcal{M} f[\psi(t)]$, where $\mathcal{M}$ is the operator that projects $f$ onto the $m$-dimensional space of eigenvectors of $\mcL$ that have eigenvalues near zero. This projection can be written formally as
\begin{equation} \label{Eq10:fsynDef}
    f_{\rm syn}(x,t) = \sum_{j=1}^m v_j(x,t)\frac{1}{L}\int_{-L/2}^{L/2} w_j^{\dagger}(x,t)f(x,t)\,dx.
\end{equation}\\
We then use $f_{\rm syn}$, rather than $f$, to propagate $\psi$. We exclude the zero eigenvalue and its eigenvector for computational convenience since its effect is to shift the entire waveform due to computational noise without affecting the interaction between its components. 

While this basic approach has been widely used in the past in analytical studies of, for example, soliton perturbation theory \cite{lamb_1980,ablowitz_1981,kaup_1990,georges_1995}, it has not previously been used in systematic computational studies.
\begin{figure}[httb]
\centering
\includegraphics[width=1\textwidth]{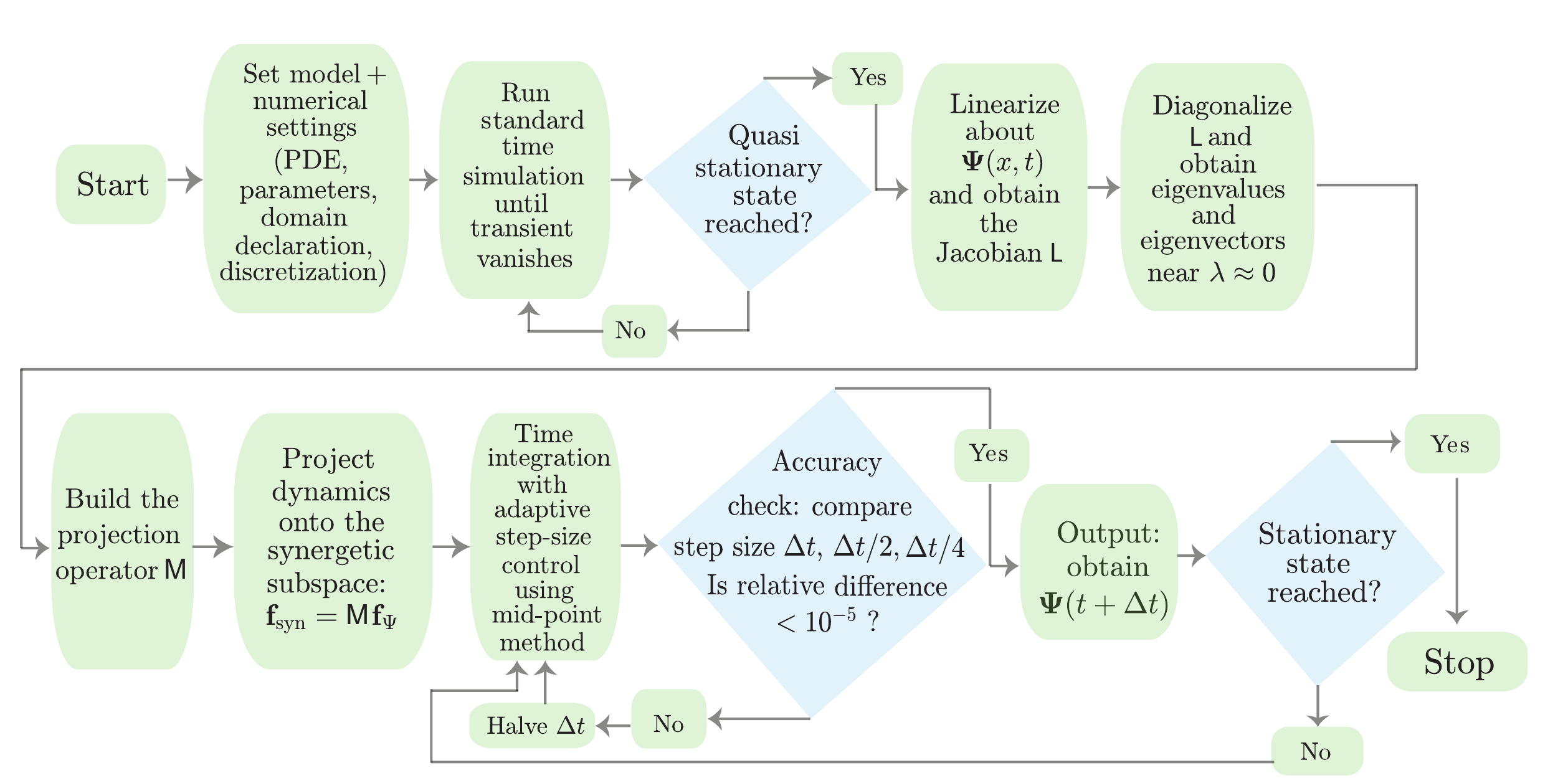} 
\vspace{-10pt}
\caption{Synergetic method algorithm flowchart}
\end{figure}

\subsection{Discretization Approach}
To make computational use of this approach, we reduce the partial differential equation (PDE) in Eq.~\ref{Eq6:LLE} to a high-order ordinary differential equation (ODE).  This discretization is conventional \cite{zhen_19,wang_2018}.  We divide the domain $[-L/2,L/2)$ into $N$ equally spaced points $\{x_j\}$.  In all the results presented here, we used $N=1,024$.  We also experimented with $N=512$ and $N=2,048$.  The former did not consistently yield sufficient accuracy and the latter led to no visible difference in our results. After discretization, both $\Psi$ and ${f}_{\Psi}$, defined in Eq.~\ref{Eq8:RLLE}, become $2N$-dimensional column vectors $\mathbf{\Psi}$ and $\mathbf{f}_{\Psi}$, and $\mcL$ becomes a $2N\times2N$ matrix $\mathsf{L}$. This discretization is straightforward with the exception of the second-order differential operator, $\partial^2/\partial x^2$ that appears in Eqs.~\ref{Eq8:RLLE} and \ref{Eq9:define_L}.  We make the replacement 
 \begin{equation} \label{Eq12:DiscDiff2}
     \frac{\partial^2}{\partial x^2} \rightarrow -\mathsf{F}_N^{-1}\mathsf{\Theta}_N\mathsf{F}_N,
 \end{equation}
where $\mathsf{F}_N$ is the $N\times N$ discrete Fourier transform matrix \cite{briggs_1995} and $\mathsf{\Theta}_N$ is the $N\times N$ diagonal matrix whose $n$-th element equals $(n-1)^2$.  While this approach produces dense $N\times N$ sub-matrices in  $\mathsf{L}$, we have verified that this approach achieves a higher combination of accuracy and efficiency than finite-difference approaches.  This approach has been used in the past to study the stability of both laser-based and microresonator-based frequency combs \cite{zhen_19,Wang_2014}.

To compute $\mathbf{f}_\textrm{syn}$ as described in Eq.~\ref{Eq10:fsynDef}, we first compute the $m$ eigenvalues $\lambda_j$ of $\mathsf{L}$ that are clustered near zero, along with the corresponding (right) eigenvectors $\mathbf{v}_j$ and the adjoint (left) eigenvectors $\mathbf{w}_j^{\dagger}$.  These eigenvectors satisfy the relations
 \begin{equation} \label{Eq13:Relns}
     \mathsf{L}\mathbf{v}_j = \lambda_j\mathbf{v}_j,\quad \mathbf{w}_j^{\dagger}\mathsf{L}=\lambda_j\mathbf{w}_j^{\dagger}, \quad \mathbf{w}_j^{\dagger}\mathbf{v}_k = \delta_{jk},\quad j=1,\ldots,m;\quad k=1,\ldots,2N,
 \end{equation}
where the $\mathbf{w}_j^{\dagger}$ are orthogonal to all the eigenvectors of $\mathsf{L}$ except $\mathbf{v}_j$. We now find that
 \begin{equation} \label{Eq14:fvsyn}
     \mathbf{f}_\textrm{syn} = \sum_{j=1}^m (\mathbf{w}_j^{\dagger}\mathbf{f}_{\Psi})\mathbf{v}_j=\sum_{j=1}^m \mathbf{v}_j\mathbf{w}_j^{\dagger}\,\mathbf{f}_{\Psi}.
 \end{equation}
We observe that the matrix $\mathsf{M}$ that corresponds to the operator $\mathcal{M}$ that was defined prior to Eq.~\ref{Eq10:fsynDef} can be written $\mathsf{M} = \sum_{j=1}^m \mathbf{v}_j\mathbf{w}_j^{\dagger}$, which is a $2N\times2N$ rank $m$ matrix.

Given $\mathbf{f}_\textrm{syn}$, we can apply any standard time discretization method.  In this work, we employed a midpoint Euler method, except as noted, verifying numerical accuracy by iteratively halving the time step that we explain in detail in the next section. With $\mathbf{f}_{\rm{syn}}$, the synergetic-midpoint method becomes,
 \begin{equation} \begin{split}
    \mathbf{\Psi} (x,t+\Delta t/2) &= \mathbf{\Psi}(x,t) + \frac{\Delta t}{2}\mathbf{f}_{\rm syn}\big [ \mathbf{\Psi} (x,t)\big],\\
    \mathbf{\Psi}(x,t+\Delta t)&=\mathbf{\Psi}(x,t) + \Delta t\, \mathbf{f}_{\rm{syn}} \big [ \mathbf{\Psi} (x,t+\Delta t/2)\big].
    \label{midpoint}
\end{split}\end{equation}
Here, we apply adaptive step-size control to select the largest $\Delta t$ permitted by the system dynamics at each step. Starting from a reasonable initial $\Delta t$, we increase the step size incrementally after each iteration and monitor the change in the norm of $\textbf{f}_{\rm{syn}}$ between successive steps. As long as this difference remains below a threshold of $10^{-3}$, we continue updating $\Delta t$ throughout the time evolution. While we used a method that only searches for a fixed number of the largest eigenvalues [MATLAB's {\tt eigs} routine \cite{MATLAB_EIGS}], the computational cost per step is substantially larger than a single step of the standard split-step algorithm. Thus, the computational efficiency of this approach depends critically on taking time steps that are many orders of magnitude larger than what the standard methods make possible. 

\subsection{Comparison to Other Stiff Equation Solution Methods}
To understand the computational efficiency of the synergetic method, we compare it against standard stiff solvers, specifically the backward differentiation formula (BDF), the Radau method (implicit Runge-Kutta), and the Rosenbrock method \cite{recipes_2007, hairer_1996, rosenbrock_1963}. These established algorithms typically achieve A-stability or L-stability through implicit integration, which permits significantly larger time steps than the explicit schemes. However, their robustness comes at a high computational cost: BDF and Radau methods require the iterative solution of coupled nonlinear equations via Newton’s method at each time step, while Rosenbrock methods require the factorization of a dense Jacobian matrix. For systems like the LLE with large discretizations ($N$), the recurring $O(N^3)$ linear algebra operations would require an unacceptably high computational cost.

By contrast, the synergetic method addresses stiffness not through implicit stabilization but through dimensional reduction based on Haken's slaving principle \cite{haken_1977}. By projecting the dynamics onto the eigenbasis of the linearized operator and systematically eliminating the fast-decaying radiation modes, the method allows a very large step size. A distinct advantage of our method is the resulting simplicity of implementation. Once the slave DOFs are removed, the reduced system is no longer stiff and can easily be propagated using explicit schemes, such as the explicit midpoint method used here. Furthermore, the reduced system can be adapted to any available standard numerical method.

The usefulness of the synergetic method depends on a set of assumptions that define its regime of validity. It assumes a clear separation of time scales and a small set of slowly growing or decaying eigenmodes that govern the long-time dynamics while all other eigenmodes rapidly decay. Its accuracy therefore depends on two conditions: The local linearization must remain valid over each time step, and the spectrum of the linearized equation must consist of a small number of eigenvalues whose magnitudes are orders of magnitude smaller than all the others. The real parts of all the other eigenvalues must be large and negative to correspond to rapidly damped DOFs. In contrast to the conventional stiff schemes discussed above, which can seamlessly transition between a range of time scales, the synergetic method requires a many-orders-of-magnitude separation between the rapidly damped DOFs and the DOFs that are retained. When the system changes rapidly, for example during breathing, collisions, or an abrupt parameter drift, frequent re-linearization is required. The projection onto the retained modes can lag the true dynamics, and the method will then become slower and less accurate than a conventional simulation with small steps. A practical remedy for sudden changes is a hybrid strategy. We monitor simple indicators of rapid evolution, such as growth of the residual, energy mismatch, and when a threshold is crossed we switch to short spans of SSFM evolution to model the fast transient. Once the system settles onto a new slowly evolving state, we recompute the local spectrum and return to the synergetic evolution. This hybrid approach preserves accuracy during periods of rapid evolution, while keeping the overall cost low in the long run.

\bigskip 

\section{Results}
All simulations reported here were performed on a workstation equipped with a 13th Gen Intel Core i7-13700 CPU (2.10 GHz) and 32 GB of RAM.
\subsection{Two-soliton system}\label{two_soliton_section}
We first investigate the simplest non-trivial soliton pattern---a two-soliton molecule. In the large $\alpha$ limit, Leshem et al. \cite{leshem_2026} previously derived an analytical expression for the relative motion of the two solitons in the potential well that is created by their mutual interaction, enabling validation of our numerical modeling and the new synergetic method that we describe here. Like Leshem et al.~\cite{leshem_2026}, we focus on the parameter choice $\alpha=1.9$ and $F=1.3$.\\
\begin{figure}[htb]
\centering
\includegraphics[width=0.95\textwidth]{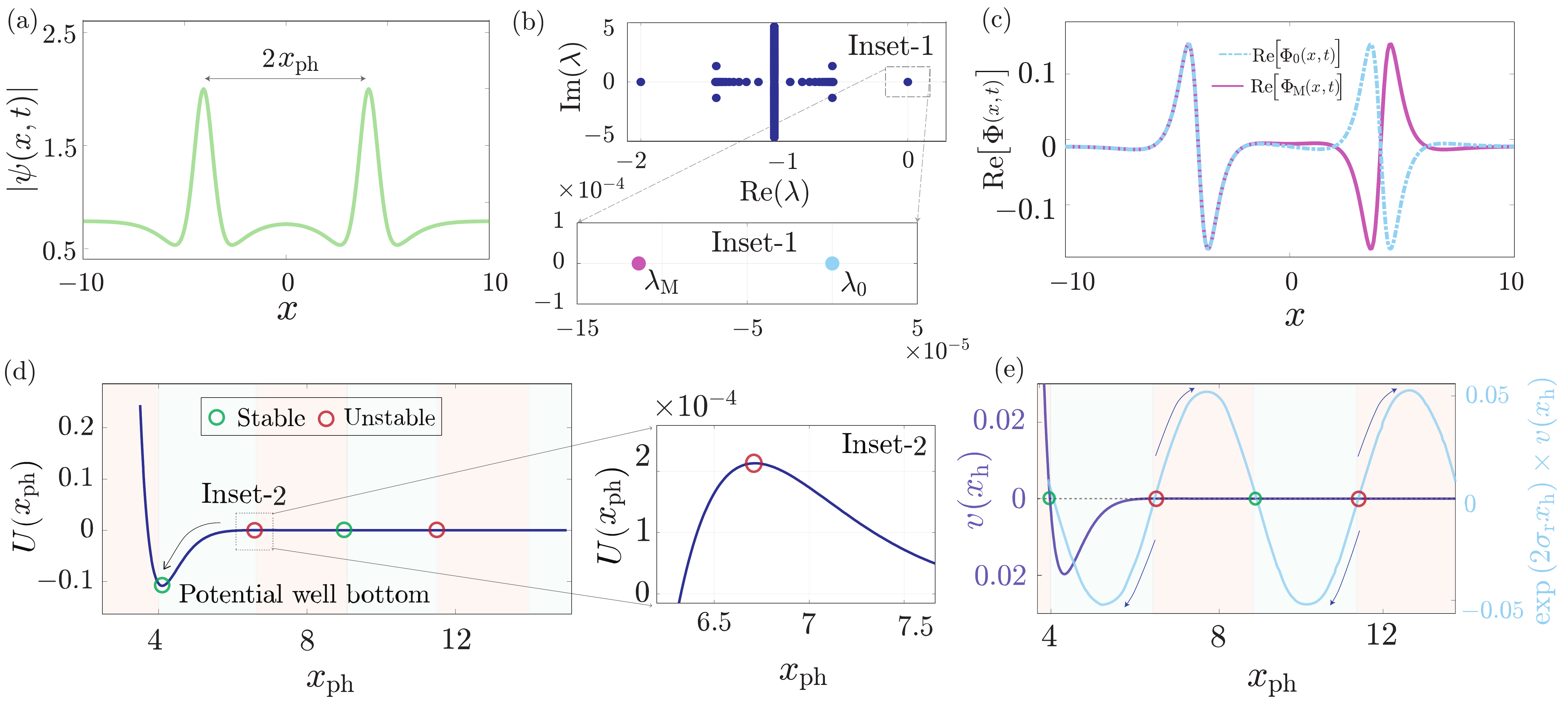}
\vspace{-10pt}
\caption{(a) Two-soliton molecule system where $x_{\mathrm{ph}}$ is half the separation distance. (b) Dynamical spectrum of the two-soliton molecule showing the presence of two eigenvalues near zero: $\lambda_{\mathrm{0}}$ corresponds to translational invariance of the soliton pair and $\lambda_{\mathrm{M}}$ is responsible for relative motion between solitons. (c) Shape of the two eigenvectors of the near-zero eigenvalues. (d) Two-soliton potential function $U(x_{\mathrm{ph}})$ as a function of half the separation distance $x_{\mathrm{ph}}$ showing the consecutive stable and unstable equilibria. Inset-2 shows $U(x_{\mathrm{ph}})$ near the potential maximum at $x_{\rm ph} = 6.7$. The potential beyond this point appears nearly flat on a linear scale due to the exponential falloff of the envelope of the oscillating potential.  (e) Soliton velocity as a function of $x_{\mathrm{ph}}$ showing the direction of movement as we start at different separations.}
\label{two_soliton_plot}
\end{figure} 

To model the interaction dynamics numerically, we consider an initial condition consisting of two pulses; however, due to the nonlinear nature of Eq.~\eqref{Eq6:LLE}, a simple superposition of two cavity solitons is not a steady-state or even a quasi-steady-state solution. Since solitons are localized, a superposition of well-separated solitons results in a quasi-steady state characterized by slightly deformed soliton waveforms. To obtain the quasi-steady-state two-pulse solution for our synergetic simulation, we start with the single soliton solution of the LLE, Eq.~\eqref{Eq6:LLE}.  This solution consists of a pulse that is approximately hyperbolic-secant-shaped $\psi_i(x)$ with a constant pedestal $\psi_c(x)$.  This solution is a stationary solution of the LLE and can be found numerically  using  MATLAB's root-finding routine {\tt FSOLVE} \cite{MATLAB_fsolve} that is based on the Levenberg-Marquardt algorithm,  as described in more detail in \cite{Wang_2014,qi_2017}. Once this solution has been found, we superpose two well-separated copies of $\psi_i(x)$ and one copy of the common pedestal $\psi_c(x)$.  We may write this solution as 
\begin{equation}
    \psi_s(x) = \psi_i(x-x_{\rm ph}) + \psi_i(x+x_{\rm ph}) + \psi_c(x),
    \label{two_soliton_start}
\end{equation}
where $x_{\mathrm{ph}}$ is half the separation distance between solitons. To obtain the quasi-steady state that we use as the starting point for the synergetic method, we can proceed in two ways, and we have used both.  The first and most direct approach is to allow the solution in Eq.~\ref{two_soliton_start} to evolve for a time $t_0$ using the SSFM until the fast transients have disappeared, but is short compared to the timescale on which the solitons interact.  In normalized units, we have found that setting $t_0$ between 100 and 200 is sufficient. In the second approach, we once again use the {\tt FSOLVE} implementation of the Levenberg-Marquardt algorithm, but we now require that $f-f_{\rm syn}=0$ rather than $f=0$.  The results are nearly identical with a fractional difference in the $L_2$ norm that is less than $10^{-5}$ in all the cases that we examined, and either approach produces the desired quasi-steady state $\psi_0(x)$, which may be written 
\begin{equation}
    \psi_\mathrm{0}(x)=\psi_{i}(x-x_{\rm ph}) + \psi_{i}(x+x_{\mathrm{ph}}) + \psi_\mathrm{c}(x) + \phi(x),
    \label{two_soliton}
\end{equation}
where $\phi(x)$ is the localized adjustment term required to form a quasi-stationary solution. The second approach is computationally faster. In Fig.~\ref{two_soliton_plot}(a), we show the two-soliton solution $\psi_\mathrm{0}(x)$ that is the starting point for our synergetic simulation.

Fig.~\ref{two_soliton_plot}(b) shows the dynamical spectrum of the two-soliton system, highlighting two eigenvalues near zero (Inset-1). The eigenvalue exactly at zero ($\lambda_0$) corresponds to the translational invariance of the soliton pair. The other eigenvalue near zero ($\lambda_{\mathrm{M}}$) corresponds to the relative motion between the solitons. This DOF governs whether the solitons move closer because they mutually attract or move apart because they mutually repel. We show the shape of the two eigenvectors corresponding to $\lambda_0$ and $\lambda_{\rm{M}}$ in Fig.~\ref{two_soliton_plot}(c).

Since we are interested in cases where the separation between solitons is much larger than their widths so that evolution is slow, it is reasonable to assume that their interaction primarily affects the soliton positions, leaving their shapes unchanged. This assumption implies that, for two interacting solitons, only the DOF associated with $\lambda_{\mathrm{M}}$ contributes to the interaction dynamics, guiding the solitons to adjust their separation and eventually reach a stationary state. This assumption is the starting point of the analytical work in \cite{leshem_2026} and is consistent with the results of the synergetic method presented here.
As was previously shown \cite{leshem_2026, synergetic_2024}, the expression for the velocity of each soliton with respect to $x_{\mathrm{ph}}$ is
\begin{equation}
    v(x_{\mathrm{ph}})=-r\exp{(-2\sigma_\mathrm{r} x)}\cos{(2\sigma_\mathrm{i} x-\phi_\mathrm{r})},
    \label{velocity}
\end{equation}
where $r=1.53$, $\sigma=\sigma_{\mathrm{r}} + i\sigma_{\mathrm{i}}=1.19-0.6i$, and $\phi_\mathrm{r}=-0.203$ for our choice $(\alpha,F) = (1.9,1.3)$. Integrating Eq.~\eqref{velocity} provides the equation for the potential well $U(x_{\mathrm{ph}})$, 
\begin{equation} 
\begin{split}
    U(x_{\mathrm{ph}})&=-\int v(x_{\mathrm{ph}}) \,dx\\ &=\frac{r\exp{(-2\sigma_\mathrm{r} x_{\mathrm{ph}})} }{2(\sigma_{\rm{r}}^2+\sigma_{\rm{i}}^2)} \Big [-\sigma_\mathrm{r}\cos{(2\sigma_\mathrm{i} x_{\mathrm{ph}}-\phi_\mathrm{r})}+\sigma_\mathrm{i} \sin{(2 \sigma_\mathrm{i} x_{\mathrm{ph}}-\phi_\mathrm{r})} \Big ].
    \label{potential_eq}
\end{split}
\end{equation}
The expression for $U(x_{\mathrm{ph}})$ in Eq.~\eqref{potential_eq} indicates that two interacting solitons experience an exponentially decaying, spatially periodic potential well as a function of $x_{\mathrm{ph}}$. 

As we can see from the plot of the potential well in Fig.~\ref{two_soliton_plot}(d), there exist alternating stable and unstable equilibrium positions as $x_{\mathrm{ph}}$ increases. Depending on their initial separation, the solitons will interact and evolve toward the nearest stable equilibrium, ultimately forming a stationary solution. The direction of motion for the interacting solitons can be clearly visualized by plotting the velocity function given in Eq.~\eqref{velocity}, as illustrated in Fig.~\ref{two_soliton_plot}(e). Inset-2 of Fig.~\ref{two_soliton_plot}(d) highlights the exponentially decaying strength of the envelope of $U(x_{\mathrm{ph}})$ near the first unstable maximum, illustrating how the potential rapidly flattens as $x_{\mathrm{ph}}$ increases and shows that soliton interactions become exponentially weaker beyond the first potential well. Hence, simulations that start with soliton separations near the second or third potential well and tracking the soliton motion as it approaches the nearest equilibrium is challenging due to step size limitations.

\begin{figure}[htb]
\centering
\includegraphics[width=1\textwidth]{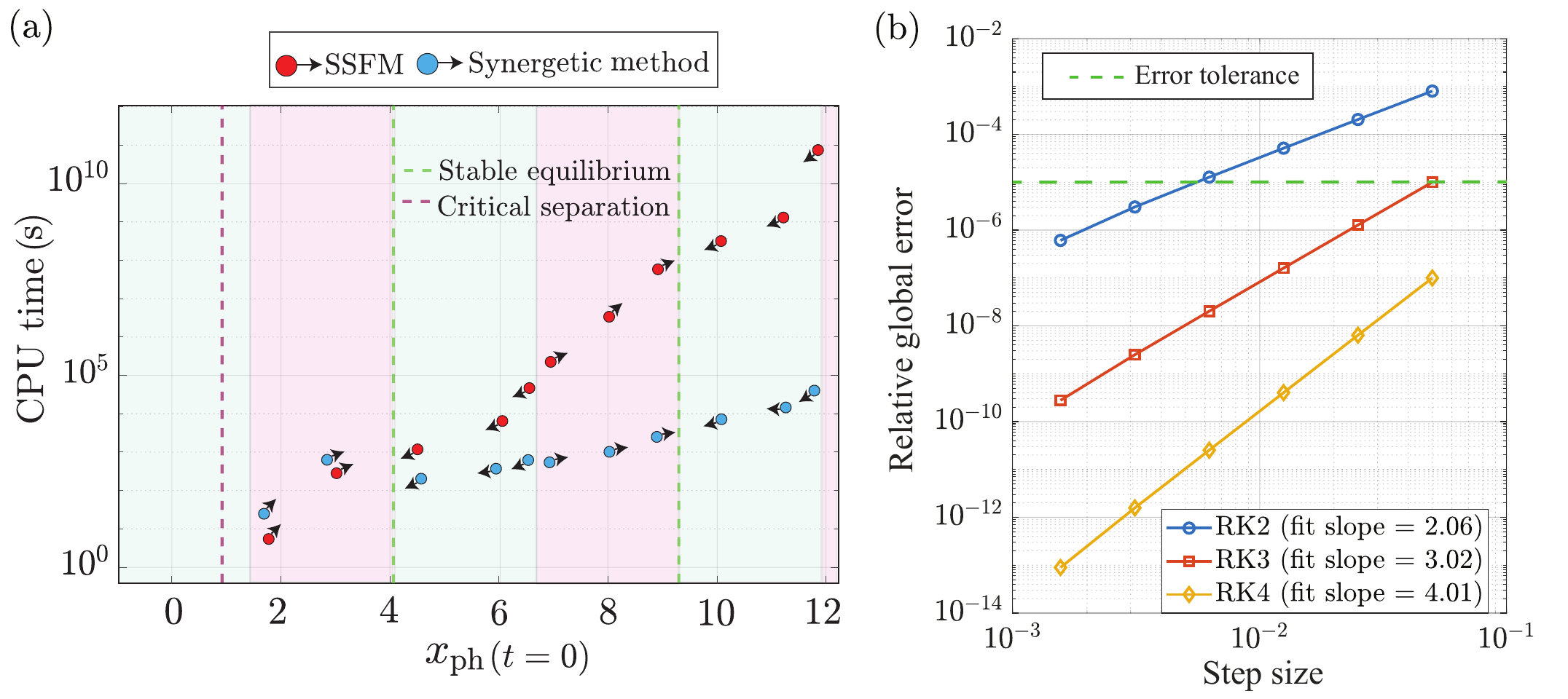} 
\vspace{-10pt}
\caption{(a) CPU time comparison between the synergetic method and the SSFM for the two-soliton system as a function of $x_{\rm{ph}}(t=0)$ as the solitons approach equilibrium starting at different separations. (b) Time comparison and error analysis for the synergetic method using different order explicit Runge-Kutta integrators; RK2 = second-order, RK3 = third-order, RK4 = fourth-order.} 
\label{algorithm_plot}
\end{figure} 

We simulate the two-soliton interaction with the synergetic method as described in Eq.~\ref{midpoint}, and also with the SSFM to make a simulation time comparison. We compare the CPU time required by the synergetic method and the SSFM to reach the nearest minimum for various initial values of $x_{\rm{ph}}$ [$x_{\rm ph}(t=0)$] as presented in Fig.~\ref{algorithm_plot}(a). For the SSFM method, the simulation time increases exponentially as solitons interact around the second well, requiring between $10^5$ and $10^{10}$ seconds of computation time (extrapolated from scaling near the well), which is infeasible in practice. On the other hand, using the synergetic method, slower soliton interactions enable significantly larger step sizes. Around the second potential well, we can utilize time steps ranging from $10^3$ to $10^6$,  accelerating the simulation of slow interactions by factors of $10^{2}$ to $10^{5}$.

We also validated the synergetic method as a numerical scheme by implementing the synergetic method with second, third, and fourth order Runge Kutta integrators (RK2, RK3, and RK4, respectively) \cite{recipes_2007} in the two-soliton case. To determine accuracy, we initialized the system at a separation of $x_{\rm{ph}}=4.5$ and tracked its evolution to the first stable equilibrium at $x_{\rm{ph}}=4$ using the solution obtained with the smallest step size as a reference, and computing the relative global error against this reference for increasingly doubled step sizes. As shown in Fig.~\ref{algorithm_plot}(b), the measured error scales as $\Delta t^{p}$ with fitted slopes of $2.06$, $3.02$, and $4.01$ for RK2, RK3, and RK4, respectively, in close agreement with the theoretically-expected $p$-th order convergence \cite{recipes_2007} of each order. The green dashed line in Fig.~\ref{algorithm_plot}(b) marks the error tolerance of $10^{-5}$ that we enforced throughout this work. In every synergetic method simulation reported in the paper, the step size was selected such that the relative global error remained below this threshold. The choice of RK2 (mid-point Euler) for the remainder of our simulations, despite the faster convergence offered by RK3 and RK4, reflects a system-dependent trade-off for the error tolerance that we selected and the range of step sizes that are practical to take. Each step of the synergetic method requires re-linearization of the system about the current state, and higher-order schemes demand multiple such linearizations per step, raising the per-step computational cost. When the rate of evolution is relatively large, limiting the step size, we found that the higher-order methods lose their advantage in our system.

We now shift our focus to more complex structures where analytical solutions are not available.

\subsection{Three-soliton molecule}
\begin{figure}[htb]
\centering
\includegraphics[width=1\textwidth]{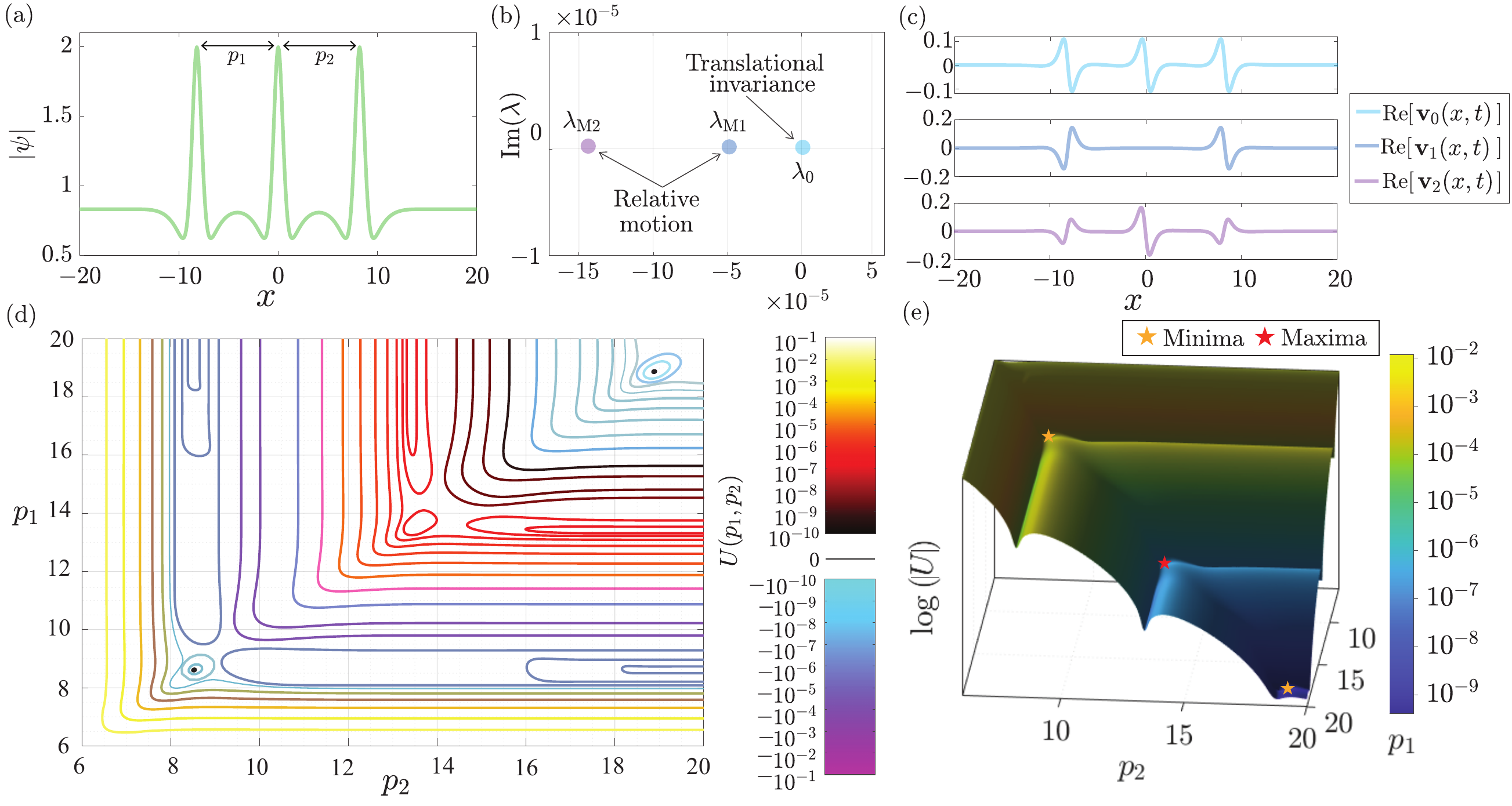} 
\vspace{-10pt}
\caption{(a) A three-soliton molecule system, where $p_1$ and $p_2$ represent the peak-to-peak separations between the left-middle and right-middle soliton pairs, respectively. (b) Three eigenvalues near or at zero from the dynamical spectrum of the three-soliton system shown in (a). The eigenvalue $\lambda_0$ is exactly at zero, corresponding to translational invariance, while $\lambda_{\rm{M1}}$ and $\lambda_{\rm{M2}}$ represent the eigenvalues associated with the relative motion between solitons. (c) The three eigenvectors, corresponding to $\lambda_0$ (top), $\lambda_{\rm M1}$ (middle), and $\lambda_{\rm M2}$ (bottom), normalized using the $L_2$ norm. (d) The two-dimensional signed contour plot of the potential function $U(p_1,p_2)$. The black dots correspond to stable equilibria. (e) The three-dimensional unsigned contour plot. }
\label{three_soliton_plot}
\end{figure}

For the three-soliton system that we show in Fig.~\ref{three_soliton_plot}(a), we begin by writing the initial condition of our synergetic simulation for different initial pulse separations as
\begin{equation}
    \psi_\mathrm{0}(x)=\psi_{i}(x) + \psi_{i}(x-p_{1}) + \psi_{i}(x+p_2) +\psi_\mathrm{c}(x) + \phi(x).
    \label{three_soliton}
\end{equation} 
This initial quasi-stationary three-soliton molecule solution is found using the same procedure that we described for the two-soliton molecule.
For simplicity, we position $\psi_i(x,t)$ at the center $(x=0)$, and define $p_1$ and $p_2$ as the peak-to-peak distances between the middle soliton and the left and right solitons, respectively. We consider the same $(\alpha, F) =(1.9,1.3)$ values that we used for the two-soliton  molecule in order to compare the two systems. 

We will begin by finding a potential function $U(p_1,p_2)$ such that 
 \begin{equation}
     v_1=\frac{d p_1 }{dt} =\dbd{U(p_1,p_2)}{p_1},\qquad v_2=\frac{d p_2}{d t} = \dbd{U(p_1,p_2)}{p_2}.
     \label{PotFunc}
 \end{equation} 
It is not evident \textit{a priori} that the velocity field $(v_1,v_2)$ is derivable from a potential. For that to hold, the Jacobian of the vector field must be symmetric, implying that it is curl-free, which we verified computationally.
This potential function describes the motion on a hyperplane in the function space of $\psi(x)$ in which $f_{\rm syn} = f$. To find this hyperplane, we start with a grid of points in the $(p_1,p_2)$, using as our initial starting point for Eq.~\eqref{three_soliton} with the unknown $\phi(x)$ set to zero.  We then force the solution onto the hyperplane by demanding that $f-f_{\rm syn} = 0$, just like when we initialize $\psi_0$ for a synergetic simulation. We selected a grid of $10^{-2}\times 10^{-2}$ equally spaced $(p_1,p_2)$ points in the range $[4,20]\times[4,20]$.  We then use one time step of the synergetic method to determine $v_1(p_1,p_2)$ and $v_2(p_1,p_2)$, the time derivatives of $p_1$ and $p_2$ on this hyperplane.  Given $v_1$ and $v_2$, we can then determine the potential function by integration, writing
\begin{equation}
    U(p_1,p_2) = -\int_{p_{1}}^\infty dp_1'\int_{p_{2}}^\infty dp_2' \big[v_1(p_1',p_2') + v_2(p_1',p_2') \big].
    \label{potentialEq}
\end{equation} 
In practice, due to the exponential falloff of the envelope of $f_{\rm syn}$, we can use any point on the outer edge of the grid ($p_1 = 20$ or $p_2 = 20$) as our point at $\infty$.  We used a trapezoidal integration scheme, which provided sufficient accuracy.  After computing the potential function on the grid, we interpolated it to produce a contour of the potential function.

We show the result of this computation in Fig.~\ref{three_soliton_plot}.  There is a stable stationary solution at the point $(p_1,p_2) = (8.2,8.2)$.  We show this soliton solution in Fig.~\ref{three_soliton_plot}(a) and its dynamical spectrum in Fig.~\ref{three_soliton_plot}(b). The 3:1 ratio of the damping eigenvalues, $\lambda_{\rm M2}$ and $\lambda_{\rm M1}$ respectively, is the same ratio that exists for three equal-mass bobs attached by two equal-strength springs, where the springs are moving in a viscous medium \cite{Cline_2021}, where it has long been known that the motion can be derived from a potential function even though the system is dissipative.  It is this observation that motivated our search for a potential function, and this result is not as surprising as it may at first appear since our system like the three-bob system is dominated by nearest-neighbor interactions. In Fig.~\ref{three_soliton_plot}(c), we show the eigenmodes that correspond to the eigenvalues at or near zero.  The eigenmode at 0 corresponds to translation, the eigenmode at $\lambda_{\rm M1}$ corresponds to symmetric stretching, and the eigenmode at $\lambda_{\rm M2}$ corresponds to anti-symmetric stretching.  In Fig.~\ref{three_soliton_plot}(d), we show a signed logarithmic contour plot of $U(p_1,p_2)$.  We show stable equilibria at $(p_1,p_2) = (8.2,8.2)$ and $(p_1,p_2) = (18.2,18.2)$ as black dots.  Along the diagonal, the potential mirrors that of a two-soliton system, not only in the positions of its minima and maxima but also in the exponential decay of its strength with increasing $p_1$ and $p_2$. Off the diagonal, where $p_1$ and $p_2$ differ significantly, another minimum appears in which one pair is at the first minimum and the other is at the second, resulting in an additional stable locking condition.  Finally, in Fig.~\ref{three_soliton_plot}(e), we show an unsigned three-dimensional contour plot of $\log(|U(p_1,p_2)|$, which clearly exhibits the overall exponential falloff as $p_1$ and $p_2$ increase.

\begin{figure}[htb]
\centering
\includegraphics[width=1\textwidth]{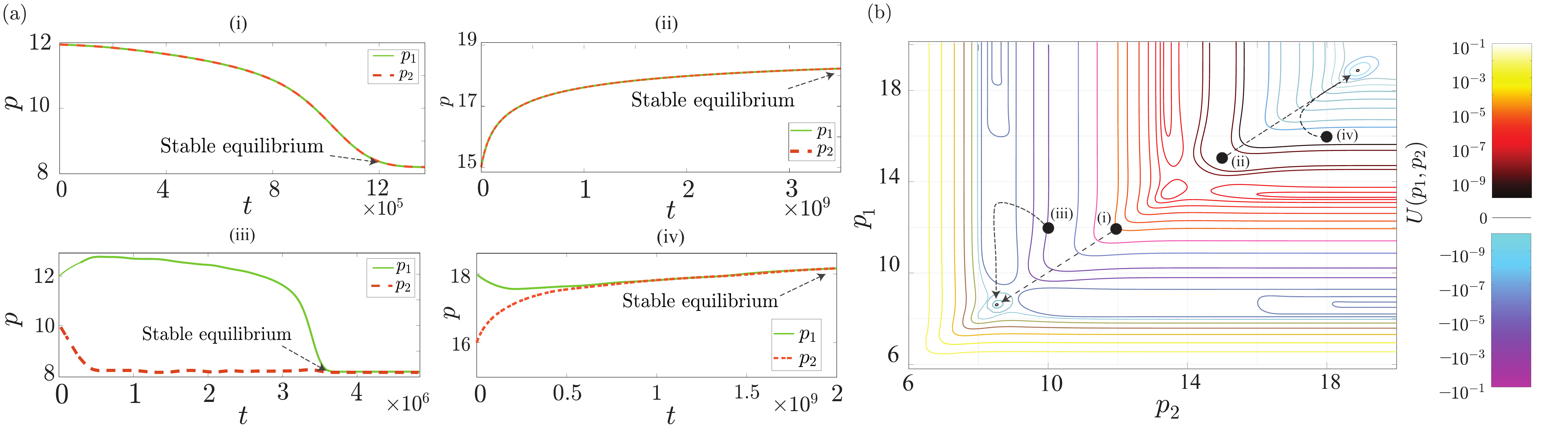} 
\vspace{-10pt}
\caption{ Illustration of the motion of the molecule trajectories governed by the three-soliton potential function. (a) Evolution of the inter-soliton separations $p_1$ and $p_2$ as functions of time, obtained using the synergetic method, for four representative initial conditions: (i) $(p_1, p_2) = (12, 12)$, (ii) $(p_1, p_2) = (15, 15)$, (iii) $(p_1, p_2) = (12, 10)$, and (iv) $(p_1, p_2) = (16, 18)$. (b) Two-dimensional signed contour map of the potential function with the four initial conditions (filled circles) and the corresponding trajectories (dashed arrows).}
\label{three_soliton_p}
\end{figure}

To verify that the potential function constructed via the point-integral approach correctly captures the dynamics of the three-soliton system and to illustrate the path of the trajectories on the $(p_1,p_2)$ hyperplane, we performed a set of evolution simulations using the synergetic method and compared their endpoints to the equilibria predicted by $U(p_1,p_2)$. We selected four initial conditions on the $(p_1,p_2)$ hyperplane to follow distinct trajectories: a symmetric point $(12, 12)$ near the diagonal close to the first minimum, a symmetric point $(15, 15)$ near the second minimum,  an asymmetric point $(12, 10)$ near the first minimum, and another asymmetric point $(16,18)$ near the second minimum. For each initial condition, we evolved the system using the synergetic method and tracked the time evolution of $p_1$ and $p_2$ as the solitons relaxed toward their stationary configurations. As shown in Fig.~\ref{three_soliton_p}(a), all four trajectories converge to the minimum predicted by the calculated potential function. This agreement is shown explicitly in Fig.~\ref{three_soliton_p}(b), where the trajectories are overlaid on the potential contour and seen to terminate at the corresponding minima. We observe that as required, the trajectories cross the contour lines at right angles. It is thus possible in principle to determine the motion of any trajectory by making use of the potential function.

In our synergetic simulations, the time was normalized to the photon lifetime.  In microresonator applications, quality factors can vary between $10^6$ and $10^9$;  typical light wavelengths are 1.0 and 1.5 $\mu$m (300 THz and 200 THz, respectively).  Thus, photon lifetimes can vary between about 500 ps and 800 ns.  The simulation over a normalized time $t=10^9$ corresponds to 0.5 s in the former case and 800 s in the latter case.  Thus, we have demonstrated that our method can describe phenomena on a laboratory timescale of fractions of a second to many seconds.

\subsection{Eight-soliton molecule}
\begin{figure}[htb]
\centering
\includegraphics[width=1\textwidth]{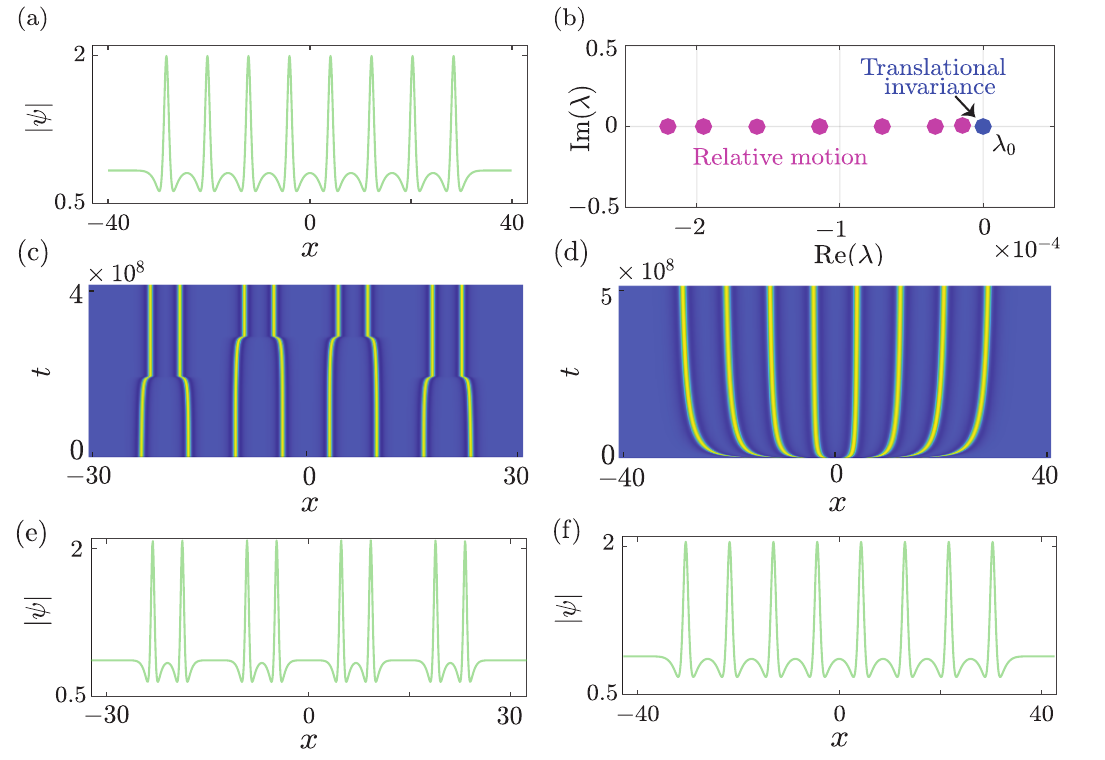} 
\vspace{-10pt}
\caption{(a) Eight soliton molecule profile. (b) Dynamical spectrum of the eight-soliton molecule, where the eigenvalue at zero corresponds to translational invariance and the remaining eigenvalues correspond to relative motion. Evolution of the eight-soliton molecule toward a stationary state, with each soliton having an initial equal peak-to-peak separation of 6.5. (d) Evolution of the eight-soliton molecule with initial peak-to-peak separations of 4. (e) and (f)  The final stationary states corresponding to (c) and (d) respectively.}
\label{eight_soliton_plot}
\end{figure}

As a final application of the synergetic method, we investigate an eight-soliton molecule as an example of complex soliton structures to test its performance in a complex system. In Fig.~\ref{eight_soliton_plot}(a), we present the eight-soliton molecule system using the same $\alpha=1.9$ and $F=1.3$ values as in the previous cases, and in Fig.~\ref{eight_soliton_plot}(b) we show the corresponding near zero eigenvalues of the system. As anticipated by the analogy to equal-mass bobs attached by equal-strength springs moving in a viscous medium, there are eight eigenvalues near zero, with one representing translational invariance and the remaining ones corresponding to relative motion. The eigenvalues are once again in the same ratios as in the analogous bob and spring system. For the eight-soliton molecule, the starting point for the synergetic simulation may be written $\psi_{0}(x)=\sum_{k=1}^{n}\psi_{i}(x-p_k)+ \psi_\mathrm{c}(x) + \phi(x)$, where $n=8$ in this case and $p_k$ denotes the displacement of the peak position from $x=0$. We initially place the pulses symmetrically around $x=0$ in the cases that we study in this section, and we find $\phi(x,t_0)$ in the same way that we did for the two- and three-molecule systems.
\\

In this system, we studied two eight-molecule systems that are displaced from equilibrium, and we observed their evolution as they approach a stationary state. In Fig.~\ref{eight_soliton_plot}(c), we start with an initial separation of 6.5.  In this case, the solitons undergo a period of slow evolution before the system relatively rapidly falls into a potential well where the solitons are unequally spaced.  In Fig.~\ref{eight_soliton_plot}(d), we start with initial separations of 4, and the solitons slowly and symmetrically evolve towards the stable stationary state in which the separation between all the solitons is $8.125$. In these two simulations, the time steps that we took varied between $\Delta t = 3\times10^4$ and $6\times10^4$.
In the microresonator application, the total time $t=5 \times 10^8 $ corresponds to laboratory times of 0.25 s to 400 s, which are well within the capability of laboratory measurement.

The two cases reported here are representative examples drawn from a rich set of molecular patterns whose evolution we modeled with the synergetic method.

\bigskip

\section{Conclusion}
In this work, inspired by Haken's slaving principle, we have introduced the synergetic method, a numerical method that is designed to efficiently model the slow evolution of complex patterns in nonlinear systems when the pattern evolution is many orders of magnitude slower than the dissipative processes in these systems, which lead in nature to the rapid disappearance of the large number of damped DOFs and the domination of the small number pattern DOFs.  As a result of the large separation in time scale between the pattern DOFs and the damped DOFs, standard numerical methods like the SSFM in the case of soliton evolution, become stiff.  In contrast to standard stiff methods, which can seamlessly transition between a range of fast and slow time scales, our method is only applicable in cases in which there is a clear and large separation between the damped and pattern DOFs.  However, this situation often occurs in nature.

In this method, we find the eigenmodes and eigenvalues of the Jacobian of the evolution operator and then project this operator onto the subspace that corresponds to the slow evolution of the patterns.
By projecting the equation of motion onto the eigenbasis of the Jacobian and retaining only the near-zero eigenmodes, the synergetic method removes the damped DOFs, whose presence constrains conventional approaches, thereby enabling time steps many orders of magnitude larger than is otherwise possible. An advantage of this method is its simplicity since it can be combined with any explicit or implicit scheme to do the time-stepping. For almost all of our evolution studies, we used an explicit midpoint Euler method with a variable step size, using the criterion that the relative change in the norm of $\textbf{f}_{\rm syn}$ is less that $10^{-3}$. We verified that this choice was sufficient by checking in selected instances that decreasing this relative error did not change the result.

 We applied this method to soliton patterns of the Lugiato-Lefever equation, which is a damped-driven nonlinear Schr\"odinger equation, and has been used to model optical patterns in a variety of settings. In recent years, it has become widely used as a starting point for modeling soliton interactions in microresonators, which is currently a topic of great scientific and technological importance.  We focused on three soliton-molecule patterns:  two-soliton molecules, three-soliton molecules, and eight-soliton molecules.
 For the two-soliton molecule, we validated the synergetic method by comparing our results to the previously derived analytical results, and we demonstrated simulation speed-ups of $10^{2}$ to $10^{5}$ relative to the SSFM.  For the three-soliton molecule, we numerically mapped the full two-dimensional inter-soliton potential function on its hyperplane, and we explicitly tracked the evolution of several initial patterns towards stable stationary states, verifying that they follow the path predicted by the gradient of the potential.  For the eight-soliton molecule, we tracked the evolution towards stationary states for two initial patterns, verifying that the synergetic method can follow the evolution of complex patterns.

 In the microresonator application, the damping time is the photon lifetime, and photon lifetimes vary between 500 ps and 800 ns.  Our method makes it possible to simulate soliton patterns in these devices on laboratory timescales as large as hundreds of seconds, which has not been feasible in the past.  We expect similar gains in the wide range of physical, chemical, and biological contexts where a similar large difference in timescales exists.

An important application of this work is to study the stability of patterns in the presence of noise and long-term environmental drift.  That is a question of scientific importance across a broad range of optical, chemical, and biological systems.  We have already begun work on this question in the context of microresonators \cite{akter_2025_noise}.

\section{Acknowledgements}

S.A., P.S., L.C., and C.M. acknowledge support from The National Center for Manufacturing Sciences (Cooperative Agreements 2022138-142232 and 2023200-142386 as subawards from DoD Cooperative Agreements HQ0034-20-2-0007 and HQ0034-24-2-0001) and from the National Institute of Standards and Technology (Grant No.~60NANB24D106). C.M. also acknowledges a useful discussion with Prof.~Xu Yi of the University of Virginia who encouraged him to address the question of how to do simulations on a laboratory timescale for microresonators.

\bibliographystyle{elsarticle-num} 
\bibliography{reference}

\end{document}